\documentclass{article}
\usepackage{spconf,amsmath,graphicx, caption}
\usepackage{xspace}
\usepackage{xcolor}
\usepackage{colortbl}
\usepackage{cellspace}
\usepackage{booktabs}
\usepackage{float}
\usepackage[section]{placeins}
\usepackage{enumitem}

\newcommand{\ignore}[1]{}

\def\webdoc{WEBDOC\xspace}

\title{Large-scale Language Model Rescoring on Long-form Data}
\name{Author(s) Name(s)}
\address{Author Affiliation(s)}

\name{
\begin{tabular}{c}
Tongzhou Chen\textsuperscript{*},
Cyril Allauzen\textsuperscript{*},
Yinghui Huang,
Daniel Park,
David Rybach,
W. Ronny Huang,\\
Rodrigo Cabrera,
Kartik Audhkhasi,
Bhuvana Ramabhadran,
Pedro J. Moreno,
Michael Riley
\end{tabular} \thanks{$^*$Equal Contribution.\newline © 2023 IEEE. Personal use of this material is permitted. Permission from IEEE must be obtained for all other uses, in any current or future media, including reprinting/republishing this material for advertising or promotional purposes, creating new collective works, for resale or redistribution to servers or lists, or reuse of any copyrighted component of this work in other works.}
}
\address{Google LLC, USA}

\begin{document}

\maketitle

 
\small
\ninept
\begin{abstract}
In this work, we study the impact of Large-scale Language Models (LLM) on Automated Speech Recognition (ASR) of YouTube videos, which we use as a source for long-form ASR. We demonstrate up to 8\% relative reduction in Word Error Eate (WER) on US English (en-us) and code-switched Indian English (en-in) long-form ASR test sets and a reduction of up to 30\% relative on Salient Term Error Rate (STER) over a strong first-pass baseline that uses a maximum-entropy based language model. Improved lattice processing that results in a lattice with a proper (non-tree) digraph topology and carrying context from the 1-best hypothesis of the previous segment(s) results in significant wins in rescoring with LLMs. 
We also find that the gains in performance from the combination of LLMs trained on vast quantities of available data (such as C4~\cite{raffel2020exploring}) and conventional neural LMs is additive and significantly outperforms a strong first-pass baseline with a maximum entropy LM.

\end{abstract}


\section{Introduction}



Large-scale language models (LLM), such as as BERT~\cite{wang2019bert}, T5~\cite{2020t5}, GPT-3~\cite{brown2020language}, and PaLM~\cite{chowdhery2022palm}, have proven to be successful in  natural language processing (NLP) tasks such as, Question Answering, Text Summarization, and other Zero Shot learning applications. These models are trained on vast amounts of text data and have yielded state-of-the-art results across several NLP and search tasks. However, there is very limited work on the use of these LLMs in Automated Speech Recognition (ASR). 

Recent research has focused on fine-tuning GPT, GPT-2 and BERT models with small amounts of in-domain data showing that they tend to outperform the performance of conventional Neural LMs such as transformer LMs trained on the same data~\cite{9688232}. The authors in ~\cite{salazar-etal-2020-masked} propose the use of pseudo-likelihood scores and show that rescoring N-best hypotheses from an ASR  model can yield significant wins on Librispeech but there is always a trade-off between in-domain modeling and fine-tuning a model trained with far more text. An alternate approach to directly predict the oracle hypothesis was originally proposed in~\cite{9054745} and  used in ~\cite{chiu2021innovative} to re-rank the  N-best hypothesis using scores from BERT. 

In this paper, we scale the use of LLMs to ASR  on YouTube videos, which we use as a source for long-form ASR. We show the importance of lattice quality and contextual augmentation for long-form ASR and compare the performance of LLMs with other neural and maximum entropy based LMs using two metrics: Word Error Rate (WER) and \textit{Salient Term Error Rate} (STER).

\section{Related Work}
\label{sec:relatedwork}
Several methods to incorporate LMs in end-to-end sequence models have been proposed in the literature. Decoding algorithms ~\cite{hori2014real, chorowski2016towards, hori2018end} employ fusion strategies, such as \textit{shallow}~\cite{peyser2020improving}, \textit{cold}~\cite{sriram2017cold}, \textit{deep}~\cite{gulcehre2015using} and \textit{component}~\cite{shan2019component} fusion. However, the wins from incorporating LMs in this fashion have been relatively small for large scale ASR~\cite{kannan2018analysis}. The Hybrid Autoregressive Transducer (HAT) model introduced in \cite{variani2020hybrid} for encoder-decoder models, allowed for the computation of an internal language model component that can be quantified and appropriately interpolated with an external language model (ELM). The density ratio method proposed in ~\cite{mcdermott2019density} offers a theoretically grounded solution to leverage an external language model while separating out the \textit{acoustic likelihood} score and the internal LM score on the source domain. This modular framework lends itself to principled approaches of LM rescoring and adaptation thus overcoming some of the shortcomings of the aforementioned LM integration strategies \cite{variani2020hybrid,allauzen21hat}.

ASR systems perform best when the training data is matched to the target domain. However, end-to-end ASR models are trained on large quantities of available speech data and the LM is trained on the limited text data available in the target domain, thus enabling cross-domain transfer. Alternatively, Large LMs are trained on vast quantities of text and subsequently fine tuned on  target domain text. In both scenarios, finding an optimal combination of the end-to-end ASR model, with its implicitly trained internal LM and the external LM, is critical for best performance in the target domain.
Neural Oracle Search leverages HAT factorization for LM rescoring with an external LM to directly pick the oracle hypothesis~\cite{9054745}, while others have explored on-device neural and biasing LM integration~\cite{sainath2021efficient} and compared rescoring and deliberation~\cite{hu2022improving}, demonstrating wins across all tasks.

In this paper, we study the impact of LLMs within the HAT framework for long-form ASR. Using data from two different sources, US English (en-us) and Indian English (en-in) which is heavily code-switched with Hindi and other Indian languages, we show that wins of up to 8\% relative can be obtained in long-form ASR  while achieving a reduction of up to 30\% relative on Salient Term Error Rate (STER) over a strong first-pass baseline that uses a maximum-entropy based language model. We also demonstrate the importance of improved lattice quality that results in a lattice with a proper (non-tree) digraph topology and carrying context from the 1-best hypothesis of the previous segment(s)  obtain best performance with LLMs.  We  find that both Text-to-Text Transfer Transformer (T5) \cite{2020t5} and its multilingual counterpart, MT5 \cite{xue-etal-2021-mt5} are complementary to conventional neural LMs and outperform a strong first-pass baseline that utilizes a maximum entropy LM.


\section{Large Language Models}
\label{sec:methods}
Several LLMs have been proposed to date with significant improvements on varied NLP tasks. In this work, we mainly focus on two LLMs, T5 and PaLM, ranging in size from 3B to 540B parameters, summarized in Table \ref{tab:lm-sizes}.  The conventional neural LM used for comparisons is a conformer LM described in Section~\ref{sec:baseline-lms} and comprising of 70M parameters.

\begin{table}[htbp]
  \centering 
  \ignore{\begin{tabular}{l|r}
    \toprule
    \textbf{Model} & \textbf{Size} & textbf{Model} & \textbf{Size}  & textbf{Model} & \textbf{Size} \\
    \midrule 
    T5~\cite{2020t5} S & 60M \\
    Neural LM & 70M \\
    T5 L & 770M \\
    T5 XL & 3B \\
    MT5~\cite{xue-etal-2021-mt5} XL & 3.7B \\
    MaxEnt LM & 4.5B \\
    PaLM~\cite{chowdhery2022palm} S & 8B \\
    T5 XXL & 11B \\
    MT5 XXL & 13B \\
    PaLM M & 62B \\
    PaLM L & 540B \\
    \bottomrule
  \end{tabular}
  {\footnotesize\begin{tabular}{lr|lr|lr}
    \toprule
    \textbf{Model} & \textbf{Size} & \textbf{Model} & \textbf{Size} & \textbf{Model} & \textbf{Size} \\
    \midrule 
    T5~\cite{2020t5} S & 60M & MT5~\cite{xue-etal-2021-mt5} XL & 3.7B & MT5 XXL & 13B \\
    Neural LM & 70M & MaxEnt LM & 4.5B & PaLM M & 62B \\
    T5 L & 770M & PaLM~\cite{chowdhery2022palm} S & 8B & PaLM L & 540B \\
    T5 XL & 3B & T5 XXL & 11B & & \\
    \bottomrule
  \end{tabular}}}
 {\footnotesize\begin{tabular}{p{0.07\textwidth}r|p{0.025\textwidth}r|p{0.04\textwidth}r|p{0.04\textwidth}r}
    \toprule
    \textbf{Conventional LMs} & \textbf{Size} & \textbf{T5}~\cite{2020t5} & \textbf{Size} & \textbf{MT5}~\cite{xue-etal-2021-mt5} & \textbf{Size} & \textbf{PaLM}~\cite{chowdhery2022palm} & \textbf{Size} \\
    \midrule 
    Neural LM & 70M & S & 60M & & & S & 8B \\
    MaxEnt & 4.5B & M & 220M & & & M & 62B  \\
     & & L & 770M & & & L & 540B \\
    & & XL & 3B & XL & 3.7B &  &  \\
    & & XXL & 11B & XXL & 13B &  &  \\
    \bottomrule
  \end{tabular}}
  \caption{Comparison of LM sizes.}
  \label{tab:lm-sizes}
\end{table}

\subsection{T5 and PaLM}
Built on an encoder-decoder transformer-based architecture, T5 optimizes the log-likelihood of the target text given input to learn a mapping from the input to target.

While T5 is pretrained on the span corruption task,  LM and Prefix LM are two fine-tuning tasks used for language modeling. The LM task predicts the target sequence with null context input while the prefix LM task randomly splits the text into two halves, using the first half as the input to predict the second half. These fine-tuning tasks enable direct computation of log-likelihood of the target text, instead of the estimation of a pseudo log-likelihood  as proposed initially in \cite{wang2019bert} for masked LMs. Thus, given a text sequence $Y$, similar to the LM task, we can compute its T5 score $S_{\text{T5}}(Y)$ by using an empty string $\epsilon$ as input and the text sequence $Y$ as target, with the following equation:
\begin{equation}
   S_{\text{T5}}(Y) = \log P_{\text{T5}}(Y | \epsilon; \Theta_{\text{T5}}).
\end{equation}
For longer sequences, we can make better use of the previous context and compute the score in a semi-autoregressive fashion. Therefore, $Y$ can be split into multiple segments $Y_1 \dots Y_S$ and the log-likelihood of the current segment can be computed using the previous segment's context:
\vspace{-0.2cm}
\begin{equation}
\label{eqn:score-prefix}
   S_{\text{T5}}(Y) = \sum_{s=1}^S \log P_{\text{T5}}(Y_s | Y_{s-1}; \Theta_{\text{T5}}),
\vspace{-0.15cm}
\end{equation}
where $Y_0$ being $\epsilon$.

PaLM is an autoregressive LM with a decoder-only architecture. Hence the score of a text sequence can be computed straight-forwardly.

\subsection{Integration with ASR Models}

In this work, we use a first-pass model based on the conformer architecture~\cite{gulati2020conformer} that uses HAT factorization~\cite{variani2020hybrid}. Not only does HAT model provide a posterior score $S_{\text{HAT}}(Y|X)$, but it also estimates the internal LM (ILM) score. As mentioned in Section~\ref{sec:relatedwork}, when interpolating an external LM during rescoring or shallow fusion, estimating and subtracting the internal LM score yields wins. Thus, inference search maximizes:
\begin{equation}
\label{eqn:score-combination}
   S(Y, X) = S_{\text{HAT}}(Y|X) - \mu S_{\text{ILM}} (Y) + \nu S_{\text{ELM}} (Y),
\end{equation}
where $\mu$ and $\nu$ are tunable hyperparameters.

\section{Experiments}
\label{sec:experiments}

\subsection{Data}
We conduct experiments with data from two language locales, en-us and en-in. The multi-domain ASR model used in this paper is trained on several thousand hours of long-form utterances derived from YouTube videos\cite{liao2013large}
and short-form utterances that are anonymized, hand-transcribed and are  representative of Google's Voice Search traffic~\cite{narayanan2019recognizing}. \ignore{YouTube data is labeled in a semi-supervised fashion as described in \cite{liao2013large}. }
The test sets contain long-form utterances derived from 30-minute-long YouTube videos. 
\ignore{In the segmented version of the test set denoted by \textit{ytseg}, a speech-silence segmenter is used to obtain shorter, approximately two-minute long segments to address memory and speed constraints during rescoring. \ignore{We also present results on \textit{ytunseg} to simulate the performance on streaming ASR.} }
We set aside a subset containing 5\% of the test utterances as the development test to tune the hyperparameters.

The pre-training corpus used to train T5 is the publicly available, Colossal Clean Crawled Corpus(C4), while MT5 is pre-trained on the multilingual variant, MC4~\cite{xue-etal-2021-mt5}. To address code-switching seen in en-in~\cite{emond2018transliteration}, text data consisting of Indian English and Hindi Wikipedia and CCNet ~\cite{wenzek2019ccnet} collectively referred to as \webdoc{}, is used. This corpus consists of 170M sentences yielding 2.9B word tokens. We use 90\% data for training and 10\% data for validation. All data in mixed writing systems is transliterated to Latin to be consistent with ASR model training data used for en-in.

\subsection{Training Large Language Models}
\label{sec:experiments-t5}

We experimented with T5 and MT5 models of sizes XL
and XXL.
Both T5 and MT5 models were pre-trained for 1M steps using the span corruption task and then fine-tuned for 100K steps using the prefix LM task on C4/MC4. 
To address the heavy code-switching prevalent in en-in and the lack of Hindi data in MC4 corpus, we fine-tune MT5 on the LM task for an additional 300k steps on the \webdoc{} corpus.

PaLM models with three different sizes were trained as described in ~\cite{chowdhery2022palm} for the en-us task. The corpus used to train these models consisted of filtered web pages, books, Wikipedia, news articles, source code, and social media conversations. We use these pre-trained models as-is with no additional fine-tuning.

\subsection{ASR Models}
\label{sec:asr-models}

We use a first-pass ASR model based on the conformer architecture~\cite{gulati2020conformer} that uses HAT factorization~\cite{variani2020hybrid}. The encoder consists of a convolution
subsampling layer and 17-layers of conformer blocks. A conformer block is composed of a feed-forward module, multi-headed self-attention with relative positional encoding module, a convolution and a final feed-forward module, stacked together. The configuration used in this work has an encoder dimension of 512, 8 attention heads, a convolution kernel size of 32 and a decoder dimension of 640 ~\cite{gulati2020conformer}. The decoder at label $y_u$ is only conditioned on the previous two labels $y_{u-1}$ and $y_{u-2}$, with their embeddings concatenated and projected \cite{botros2021tied}. The models are trained on 80-dimensional log-mel filter bank coefficients and predict word-piece targets (4096 for en-us and 8192 for en-in).  The choice of these  parameters was determined by sweeping for best performance within the expected model size.

\subsection{Neural and Maximum-Entropy based Language Models}
\label{sec:baseline-lms}

In order to better understand the value of LLMs in ASR, we trained two state-of-the-art LMs, a conventional neural LM and a  Maximum Entropy based LM. The conventional Neural LM is a small, unidirectional, conformer LM (CLM) with 70M parameters, originally designed for on-device rescoring~\cite{sainath2021efficient}. It consists of 12 causal conformer layers, each with a dimension of 384, a feedforward layer dimension of 2048, a convolution kernel of size 15. We use 4-headed self attention with a left context size 31. The model is trained  on the same data as the LLMs to predict the same word-piece targets as the first-pass ASR model. Thus, for en-us, we trained it on C4 and for en-in, we trained it on \webdoc{} to match the fine-tuning corpus of MT5. The Maximum Entropy based (MaxEnt) LM~\cite{biadsy2014backoff,biadsy2017effectively}  is a log linear model based on N-gram and skip-gram word contexts, with a size of 4.5B parameters and is comparable to the size of the T5/MT5 XL models. It is also trained on the same data as the conventional Neural LM.


\subsection{Decoding and Rescoring}
\label{subsec:decoding-and-rescoring}
Decoding is performed by a time-synchronous beam search using the breadth-search expansion strategy \cite{tripathi19breadthsearch} where the number of active hypotheses at each frame is bounded by a beam size $k$.
A VAD-based segmenter \cite{zazo16vad} runs in parallel to the beam-search decoder. When the decoder receives an end-of-segment signal from the segmenter, a segment lattice is generated from the currently active hypotheses. If present, a rescoring LM is applied to this segment lattice, with the 1-best hypotheses from previous segments optionally provided as context. Only the best hypothesis in the lattice (eventually after rescoring) is carried forward in the beam-search for the  next segment. The final utterance lattice is obtained by concatenating all the segment lattices.

When using an ASR model with unlimited label context, each hypothesis within the beam encodes the full history from the beginning of the utterance. Hence, the segment lattice is a trie with a total number of paths (e.g. hypotheses) bounded by the beam size $k$. 

When using an ASR model where the label context is bound by $n$ \cite{pabhavalkar21statemerging}, beam-search hypotheses sharing the same label context of length $n$ will correspond to the same state in the segment lattice. This results in lattice with a proper (non-tree) digraph topology where the number of paths can grow up to exponentially in the number of states. This was shown to lead to a significant improvement in lattice quality: lattice diversity improvement and oracle WER reduction \cite{pabhavalkar21statemerging}.

The ASR models described in section~\ref{sec:asr-models} used limited label context with $n=2$. However when combining these models with the conformer LMs from section~\ref{sec:baseline-lms} during the beam search using HAT fusion results in dramatic increase of the label context limit making the resulting combined model to effectively have unlimited label context.

\section{Results}

\subsection{Lattice Quality}

\begin{table}[t]
    \centering
    \footnotesize
    \begin{tabular}{l|r@{\hspace{0ex}}r|r@{\hspace{1ex}}r|r@{\hspace{0ex}}r}
    \toprule
    \textbf{dev} & \multicolumn{2}{c|}{\textbf{Oracle WER}} &\multicolumn{2}{c|}{\textbf{WER}} &\multicolumn{2}{c}{\textbf{\#paths/segment}} \\
     & $\textbf{en-us}$ & $\textbf{en-in}$& $\textbf{en-us}$ & $\textbf{en-in}$& $\textbf{en-us}$ & $\textbf{en-in}$\\
    \midrule
    Baseline           & 7.3 & 12.8 & 12.2 & 17.2 & 4e20 & 4e13 \\
    No state merging   & 8.8 & 13.1 & 12.2 & 17.2 & 5.7 & 5.8 \\
    Neural LM fusion   & 8.4 & 11.0 & 11.6 & 15.6 & 5.2 & 5.7\\
    \bottomrule
    \end{tabular}
    \caption{Lattice quality on the en-us and en-in dev sets.}
    \label{tab:lattice-quality}
    \label{tab:en-us-lattice}
    \label{tab:en-in-lattice}
\end{table}

The success of a rescoring approach crucially depends on the quality of the hypotheses of the first-pass beam-search decoder. To assess the lattice quality, we computed metrics such as the $N$-best oracle WER and the average number of paths/hypotheses per segment for our baseline systems on the en-us and en-in development sets as reported in Table~\ref{tab:lattice-quality}.

As the contribution to first-pass model's posterior and internal LM at label $y_u$ depends only on the previous two labels, our baseline systems can leverage the state merging benefits of limited context models described in Section~\ref{subsec:decoding-and-rescoring} as demonstrated by the relatively low oracle WER and high number of paths per segments.

Lattice quality can be improved by improving first-pass modeling by integrating a neural LM in the beam-search decoding using HAT fusion. Table~\ref{tab:lattice-quality} shows this results in a significant improvement in 1-best WER. However, this causes the loss of the state merging benefits and results in an increase of oracle WER in en-us. However, this is still an significant  improvement compared to disabling state merging in the baseline systems.

\ignore{
\begin{table}  [htbp]
  \centering \footnotesize
  \begin{tabular}{l|c|c}%
    \toprule
    \textbf{en-us dev}
     & {\textbf{Before State Merging}} & {\textbf{After State Merging}}\\
    \midrule 
    \# paths per-segment        & 5.7   & 4e20\\
    \midrule 
    WER                         & 12.2  & 12.2\\
    Oracle WER                  & 8.8   & 7.3  \\
    + T5 XL rescoring           & 11.7  & 11.5\\
    + T5 XXL rescoring          & 11.6  & \textbf{11.3}\\
    \bottomrule
  \end{tabular}
  \caption{Lattice quality before and after state merging on en-us dev set.}
  \vspace{-0.5cm}
  \label{tab:en-us-lattice}
\end{table}
}

\subsection{Comparison of LMs}

In this Section, we consider the impact of LM integration on the en-us task.
Table ~\ref{tab:en-us-context} demonstrates the value of providing longer context to Large LMs. Each row contains the result of rescoring with the T5 XXL model when carrying over contexts of different lengths, i.e., of carrying over the 1-best hypotheses from different number of previous segments. We observe that carrying over previous context outperforms no context. However, longer contexts do not seem to provide additional wins. The rest of this paper thus uses contextual information from just the previous segment.

\begin{table}  [htbp]
  \centering
  \footnotesize
  \begin{tabular}{l|c}%
    \toprule
    {\textbf{WER}} & {\textbf{dev}} \\
    \midrule 
    Baseline                           & 12.2  \\
    + T5 rescoring, carrying 0 segment  & 11.6  \\
    + T5 rescoring, carrying 1 segment   & \textbf{11.5} \\
    + T5 rescoring, carrying 2 segments  & \textbf{11.5} \\
    \bottomrule 
  \end{tabular}
  \caption{WER comparison on the en-us test set for different lengths of carried over context}
  \label{tab:en-us-context}
\end{table}

Table~\ref{tab:en-us-full} presents the rescoring and fusion results on the en-us development and evaluation test sets for various LMs. First we observe that a small Neural LM edges out over the performance of a Maxent LM. Moreover, though the T5 S model, whose size is slightly smaller than the NLM, was slightly behind NLM, increasing the size of T5 leads to better results. It is also interesting to note that the NLM and T5 XXL models are complementary, as fusion can give a better 1-best WER. In addition, we experimented with more enormous PaLM LMs and they are able to brings the power of larger capacity and large amounts of training text, yielding better results than T5.

\begin{table}  [htbp]
  \centering
  \footnotesize
  \begin{tabular}{l|c|c}%
    \toprule
    {\textbf{WER}} & {\textbf{dev}} & {\textbf{eval}}\\
    \midrule 
    Baseline                    & 12.2  & 16.1\\
    + MaxEnt rescoring          & 12.2  & 16.4\\
    + NLM rescoring             & 11.8  & 15.8\\
    + T5 S rescoring            & 11.9  & 15.9 \\
    + T5 M rescoring            & 11.7  & 15.8 \\
    + T5 XL rescoring           & 11.6  & 15.7 \\
    + T5 XXL rescoring          & 11.5  & 15.7\\
    + PaLM S rescoring          & 11.5  & 15.5 \\
    + PaLM M rescoring          & \textbf{11.3}  & \textbf{15.4} \\
    + PaLM L rescoring          & \textbf{11.3}  &  -  \\
    + NLM fusion                & 11.6  & 15.6\\
    + NLM fusion \& T5 XXL rescoring & 11.4 & 15.5 \\
    \bottomrule
  \end{tabular}
  \caption{en-us WER comparison between T5 and other LMs}
  \vspace{-0.2cm}
  \label{tab:en-us-full}
\end{table}

\subsection{Code-switching Task}

In this Section, we present the performance of LLMs on a more challenging en-in task dominated by heavy code-switching.

Although MT5 is meant to be a multilingual LM, the amount of training data from the different languages is unbalanced. The training data consists of 5.67\% English, but only 1.21\% is Hindi in the Devanagari script~\cite{xue-etal-2021-mt5}. 
This imbalance between en-in and Hindi fails to capture the frequent code switches between English and Hindi predominant in the en-in test sets. To address this issue, we finetune both XL and XXL MT5 models on the \webdoc{} corpra with the LM task. We evaluate the raw MT5 model and these fine-tuned models on the en-in development set to study the effect of fine-tuning. These results are tabulated in Table~\ref{tab:en-in-fine-tune}.

\begin{table}  [htbp]
  \centering
  \footnotesize
  \begin{tabular}{l|c|c}%
    \toprule
    {\textbf{en-in dev}} & {\textbf{MT5 XL}} & {\textbf{MT5 XXL}}\\
    \midrule 
    Baseline                  & \multicolumn{2}{c}{17.2}\\
    Raw                       & 16.6  & 16.8\\
    Fine-tuned                & \textbf{16.1}  & 16.3\\
    \bottomrule
  \end{tabular}
  \caption{WER comparison on en-in dev set with raw and fine-tuned MT5 models of sizes XL and XXL}
  \vspace{-0.2cm}
  \label{tab:en-in-fine-tune}
\end{table}

It can be seen that rescoring with the fine tuned models outperforms rescoring with the raw MT5 model. This can be attributed to the lack of sufficient Hindi data in the MC4 corpus which can be fixed with data balanced fine-tuning. When compared to en-us, the wins from LLMs on en-in are less. We hypothesize that this could be related to the small size of the \webdoc{} corpus compared to MC4, in line with the data-hungry nature of LLMs~\cite{kaplan2020scaling, hoffmann2022training}. 

\subsection{Comparison of LMs on the code-switching task}

\begin{table}  [htbp]
  \centering \footnotesize
  \begin{tabular}{l|c|c}%
    \toprule
    {\textbf{WER}} & {\textbf{dev}} & {\textbf{eval}}\\
    \midrule 
    Baseline                    & 17.2 & 16.4  \\
    + MaxEnt rescoring          & 16.5 & 15.9  \\
    + NLM rescoring             & 16.2 & 15.4  \\
    + MT5 XL rescoring          & 16.1 & 15.2  \\
    + NLM fusion                & 15.6 & 15.0  \\
    + NLM fusion \& MT5 XL rescoring & \textbf{15.4} & \textbf{14.6}  \\
    \bottomrule
  \end{tabular}
  \caption{en-in WER comparison between MT5 and other LMs}
  \label{tab:en-in-full}
\end{table}

Table~\ref{tab:en-in-full} presents the rescoring results from various LMs. The MT5 XL  model is the best performing model with a WER reduction of 7.3\% relative on the evaluation test set.  On the other hand, the Conformer LM when used in shallow fusion in the first-pass shows additional wins. Since we fine-tuned MT5  on the same training data as Conformer LM, we also report the perplexity of MT5 and Conformer LM on the 10\% validation part of \webdoc{}. MT5 has a log perplexity per word of 4.15, slightly higher than the Conformer LM at 2.98 and MaxEnt at 3.69.

We observe that the Conformer LM and MT5 are complementary and the combination results in a best WER reduction of 8\% relative.

\section{Error Analysis}

To analyze the effectiveness of large LM, we select unigrams and bigrams with the highest Term Frequency Inverse Document Frequency (TF-IDF) values from the evaluation test sets (\textit{salient terms}) for the two languages studied in this paper. In general, such terms capture the topic presented in the video. On the one hand, they are important for indexing or information retrieval; on the other hand, they are more difficult to be recognized compared to frequently occurring function words (such as, "the", "of", etc.). We analyzed the performance of the baseline and the various large LMs on these \textit{salient terms} to study the impact on rare words. The \textit{Salient Term Error Rate (STER)} is reported in Table ~\ref{tab:salient}, defined as the number of deletion and substitution errors on the \textit{salient terms} divided by the total number of \textit{salient terms}. Out of a total of 600K words, approximately, 10\% words are tagged as \textit{salient terms} for en-in and 5\% for en-us. First we observe that almost all rescoring and fusion can reduce the error made on these \textit{salient terms}. In en-us, as reflected by the WER reported in Table \ref{tab:en-us-full}, T5 outperforms other LMs. In en-in, however, NLM fusion in the first pass has a bigger impact on the \textit{salient terms} than any rescoring method similar to what has been reported in~\cite{ravi2020improving}. Although MT5 has been fine tuned on the same data as the NLM, we find that it is less impactful by itself on the \textit{salient terms} in en-in. 

Although MT5 has been fine tuned on the same data as the neural LM, we find that it is less impactful by itself on the \textit{salient terms}. However, in both languages, the combination of these two LMs through interpolation is additive (last row in Table ~\ref{tab:en-in-full}) resulting in the best performance. As noted in~\cite{kaplan2020scaling, hoffmann2022training} scaling to larger and larger datasets is only beneficial when the data is high-quality and larger models require larger data sets.  This can explain some of the differences seen between these two relatively high resource languages.

\begin{table}  [htbp]
  \centering \footnotesize
  \begin{tabular}{l|c|c}%
    \toprule
     {\textbf{STER}} & {\textbf{en-us}} & {\textbf{en-in}}\\
    \midrule 
    Baseline                    & 28.8 & 20.0\\
    + MaxEnt rescoring          & 28.8 & 17.4\\
    + NLM rescoring             & 27.4 & 16.7 \\
    + T5/MT5 rescoring          & 26.7 & 17.6 \\
    + NLM fusion                & 27.2 & 15.4 \\
    + NLM fusion \& T5/MT5 rescoring & \textbf{26.4} & \textbf{12.1} \\
    \bottomrule
  \end{tabular}
  \caption{Errors analysis on \textit{salient terms} of en-us and en-in. }
  \vspace{-0.6cm}
  \label{tab:salient}
\end{table}

\section{Conclusion}

In this study, we presented the impact of LLMs (up to 350B parameters) on long-form ASR. We demonstrated up to 8\% relative reduction in Word Error Rate (WER) on US English (en-us) and code-switched Indian English (en-in) long-form ASR test sets and a reduction of up to 30\% relative on Salient Term Error Rate (STER) over a strong first-pass baseline that uses a maximum-entropy based language model. We also find that the gains in performance from the combination of LLMs trained on vast quantities of available data (such as C4~\cite{raffel2020exploring}) and conventional neural LMs is additive and significantly outperforms a strong first-pass baseline with a maximum entropy LM. To the best of our knowledge, this is the first study that scales LLMs to long-form ASR.

\bibliographystyle{IEEEtran}
\ninept
\bibliography{references}

\begin{thebibliography}{10}
\providecommand{\url}[1]{#1}
\csname url@samestyle\endcsname
\providecommand{\newblock}{\relax}
\providecommand{\bibinfo}[2]{#2}
\providecommand{\BIBentrySTDinterwordspacing}{\spaceskip=0pt\relax}
\providecommand{\BIBentryALTinterwordstretchfactor}{4}
\providecommand{\BIBentryALTinterwordspacing}{\spaceskip=\fontdimen2\font plus
\BIBentryALTinterwordstretchfactor\fontdimen3\font minus
  \fontdimen4\font\relax}
\providecommand{\BIBforeignlanguage}[2]{{%
\expandafter\ifx\csname l@#1\endcsname\relax
\typeout{** WARNING: IEEEtran.bst: No hyphenation pattern has been}%
\typeout{** loaded for the language `#1'. Using the pattern for}%
\typeout{** the default language instead.}%
\else
\language=\csname l@#1\endcsname
\fi
#2}}
\providecommand{\BIBdecl}{\relax}
\BIBdecl

\bibitem{raffel2020exploring}
C.~Raffel \emph{et~al.}, ``Exploring the limits of transfer learning with a
  unified text-to-text transformer.'' \emph{J. Mach. Learn. Res.}, vol.~21, no.
  140, pp. 1--67, 2020.

\bibitem{wang2019bert}
A.~Wang and K.~Cho, ``Bert has a mouth, and it must speak: Bert as a markov
  random field language model,'' \emph{arXiv preprint arXiv:1902.04094}, 2019.

\bibitem{2020t5}
C.~Raffel \emph{et~al.}, ``Exploring the limits of transfer learning with a
  unified text-to-text transformer,'' \emph{Journal of Machine Learning
  Research}, vol.~21, no. 140, pp. 1--67, 2020.

\bibitem{brown2020language}
T.~Brown \emph{et~al.}, ``Language models are few-shot learners,''
  \emph{Advances in neural information processing systems}, vol.~33, pp.
  1877--1901, 2020.

\bibitem{chowdhery2022palm}
A.~Chowdhery \emph{et~al.}, ``Palm: Scaling language modeling with pathways,''
  \emph{arXiv preprint arXiv:2204.02311}, 2022.

\bibitem{9688232}
X.~Zheng, C.~Zhang, and P.~C. Woodland, ``Adapting gpt, gpt-2 and bert language
  models for speech recognition,'' in \emph{2021 IEEE ASRU}, 2021, pp.
  162--168.

\bibitem{salazar-etal-2020-masked}
J.~Salazar, D.~Liang, T.~Q. Nguyen, and K.~Kirchhoff, ``Masked language model
  scoring,'' in \emph{2020 ACL}, Jul. 2020.

\bibitem{9054745}
E.~Variani \emph{et~al.}, ``Neural oracle search on n-best hypotheses,'' in
  \emph{ICASSP}, 2020, pp. 7824--7828.

\bibitem{chiu2021innovative}
S.-H. Chiu and B.~Chen, ``Innovative bert-based reranking language models for
  speech recognition,'' in \emph{SLT}.\hskip 1em plus 0.5em minus 0.4em\relax
  IEEE, 2021, pp. 266--271.

\bibitem{hori2014real}
T.~Hori, Y.~Kubo, and A.~Nakamura, ``Real-time one-pass decoding with recurrent
  neural network language model for speech recognition,'' in
  \emph{ICASSP}.\hskip 1em plus 0.5em minus 0.4em\relax IEEE, 2014, pp.
  6364--6368.

\bibitem{chorowski2016towards}
J.~Chorowski and N.~Jaitly, ``Towards better decoding and language model
  integration in sequence to sequence models,'' \emph{arXiv preprint
  arXiv:1612.02695}, 2016.

\bibitem{hori2018end}
T.~Hori, J.~Cho, and S.~Watanabe, ``End-to-end speech recognition with
  word-based rnn language models,'' in \emph{SLT}.\hskip 1em plus 0.5em minus
  0.4em\relax IEEE, 2018, pp. 389--396.

\bibitem{peyser2020improving}
C.~Peyser \emph{et~al.}, ``Improving tail performance of a deliberation e2e asr
  model using a large text corpus,'' \emph{arXiv preprint arXiv:2008.10491},
  2020.

\bibitem{sriram2017cold}
A.~Sriram, H.~Jun, S.~Satheesh, and A.~Coates, ``Cold fusion: Training seq2seq
  models together with language models,'' \emph{arXiv preprint
  arXiv:1708.06426}, 2017.

\bibitem{gulcehre2015using}
C.~Gulcehre \emph{et~al.}, ``On using monolingual corpora in neural machine
  translation,'' \emph{arXiv preprint arXiv:1503.03535}, 2015.

\bibitem{shan2019component}
C.~Shan \emph{et~al.}, ``Component fusion: Learning replaceable language model
  component for end-to-end speech recognition system,'' in \emph{ICASSP}.\hskip
  1em plus 0.5em minus 0.4em\relax IEEE, 2019, pp. 5361--5635.

\bibitem{kannan2018analysis}
A.~Kannan \emph{et~al.}, ``An analysis of incorporating an external language
  model into a sequence-to-sequence model,'' in \emph{ICASSP}, 2018, pp.
  1--5828.

\bibitem{variani2020hybrid}
E.~Variani, D.~Rybach, C.~Allauzen, and M.~Riley, ``Hybrid autoregressive
  transducer (hat),'' in \emph{ICASSP}, 2020, pp. 6139--6143.

\bibitem{mcdermott2019density}
E.~McDermott, H.~Sak, and E.~Variani, ``A density ratio approach to language
  model fusion in end-to-end automatic speech recognition,'' in \emph{ASRU},
  2019, pp. 434--441.

\bibitem{allauzen21hat}
C.~Allauzen, E.~Variani, M.~Riley, D.~Rybach, and H.~Zhang, ``A hybrid
  seq-2-seq {ASR} design for on-device and server applications,'' in
  \emph{Interspeech 2021}, 2021, pp. 4044--4048.

\bibitem{sainath2021efficient}
T.~N. Sainath \emph{et~al.}, ``An efficient streaming non-recurrent on-device
  end-to-end model with improvements to rare-word modeling,'' in
  \emph{Interspeech}, 2021, pp. 1777--1781.

\bibitem{hu2022improving}
K.~Hu \emph{et~al.}, ``Improving deliberation by text-only and semi-supervised
  training,'' \emph{arXiv preprint arXiv:2206.14716}, 2022.

\bibitem{xue-etal-2021-mt5}
L.~Xue \emph{et~al.}, ``m{T}5: A massively multilingual pre-trained
  text-to-text transformer,'' in \emph{2021 NAACL: Human Language
  Technologies}, Jun. 2021.

\bibitem{gulati2020conformer}
A.~Gulati \emph{et~al.}, ``Conformer: Convolution-augmented transformer for
  speech recognition,'' \emph{arXiv preprint arXiv:2005.08100}, 2020.

\bibitem{liao2013large}
H.~Liao, E.~McDermott, and A.~Senior, ``Large scale deep neural network
  acoustic modeling with semi-supervised training data for youtube video
  transcription,'' in \emph{ASRU}, 2013, pp. 368--373.

\bibitem{narayanan2019recognizing}
A.~Narayanan \emph{et~al.}, ``Recognizing long-form speech using streaming
  end-to-end models,'' in \emph{ASRU}, 2019, pp. 920--927.

\bibitem{emond2018transliteration}
J.~Emond, B.~Ramabhadran, B.~Roark, P.~Moreno, and M.~Ma, ``Transliteration
  based approaches to improve code-switched speech recognition performance,''
  in \emph{SLT}.\hskip 1em plus 0.5em minus 0.4em\relax IEEE, 2018, pp.
  448--455.

\bibitem{wenzek2019ccnet}
G.~Wenzek \emph{et~al.}, ``Ccnet: Extracting high quality monolingual datasets
  from web crawl data,'' \emph{arXiv preprint arXiv:1911.00359}, 2019.

\bibitem{botros2021tied}
R.~Botros \emph{et~al.}, ``Tied \& reduced rnn-t decoder,'' \emph{arXiv
  preprint arXiv:2109.07513}, 2021.

\bibitem{biadsy2014backoff}
F.~Biadsy, K.~Hall, P.~Moreno, and B.~Roark, ``Backoff inspired features for
  maximum entropy language models,'' 2014.

\bibitem{biadsy2017effectively}
F.~Biadsy, M.~Ghodsi, and D.~Caseiro, ``Effectively building tera scale maxent
  language models incorporating non-linguistic signals,'' 2017.

\bibitem{tripathi19breadthsearch}
A.~Tripathi, H.~Lu, H.~Sak, and H.~Soltau, ``Monotonic recurrent neural network
  transducer and decoding strategies,'' in \emph{ASRU}, 2019, pp. 944–--948.

\bibitem{zazo16vad}
R.~Zazo, T.~N. Sainath, G.~Simko, and C.~Parada, ``Feature learning with
  raw-waveform cldnns for voice activity detection,'' in \emph{Interspeech},
  2016, pp. 3668--3672.

\bibitem{pabhavalkar21statemerging}
R.~Prabhavalkar \emph{et~al.}, ``Less is more: Improved rnn-t decoding using
  limited label context and path merging,'' in \emph{ICASSP}, 2021, pp.
  5659--5663.

\bibitem{kaplan2020scaling}
J.~Kaplan \emph{et~al.}, ``Scaling laws for neural language models,''
  \emph{arXiv preprint arXiv:2001.08361}, 2020.

\bibitem{hoffmann2022training}
J.~Hoffmann \emph{et~al.}, ``Training compute-optimal large language models,''
  \emph{arXiv preprint arXiv:2203.15556}, 2022.

\bibitem{ravi2020improving}
V.~Ravi \emph{et~al.}, ``Improving accuracy of rare words for rnn-transducer
  through unigram shallow fusion,'' \emph{arXiv preprint arXiv:2012.00133},
  2020.

\end{thebibliography}

\end{document}